\documentclass[aps, prl,twocolumn, superscriptaddress, preprintnumbers, floatfix, longbibliography, 10pt]{revtex4-2}
\usepackage{graphicx}
\usepackage{amsmath}
\usepackage{yhmath}
\usepackage{bm}
\usepackage{xcolor}
\usepackage{epstopdf}
\usepackage{subfigure}
\usepackage{hyperref}
\usepackage[T1]{fontenc}
\usepackage{lmodern}

\begin{document}

\title{Arrays of one-dimensional conducting channels in minimally twisted bilayer graphene}

\author{Zhe Hou}
\affiliation{School of Physics and Technology, Nanjing Normal University, Nanjing 210023, China}
\author{Kai Yuan}
\affiliation{United Microelectronics Center Company Limited, Chongqing 400030, China.}
\author{Hua Jiang}
\affiliation{Interdisciplinary Center for Theoretical Physics and Information Sciences (ICTPIS), Fudan University, Shanghai 200433, China.}
\affiliation{Institute for Nanoelectronic Devices and Quantum Computing and State Key Laboratory of Surface Physics, Fudan University, Shanghai 200433, China.}

\begin{abstract}
Minimally twisted bilayer graphene (TBG) with interlayer potential asymmetry host one-dimensional (1D) topological helical states (THSs) at domain walls between AB/BA stacking regions. However, the nature of THS propagation remains elusive. Although it is widely believed that they form a two-dimensional (2D) triangular network, a few argue that they self-organize into 1D topological zigzag modes (TZMs) that propagate independently. In this Letter, we propose a protocol based on a two-terminal TBG nanoflake transport device and resolve this issue. Through rigorous calculations on the differential conductance and the nonequilibrium local density of states, we show that, these THSs indeed self-construct the 1D distorted TZMs, each bypassing the AA-stacking spots and propagating independently. By considering a long TBG nanoflake, we obtain a nearly quantized conductance plateau with its value close to 1, 2, and 3 (in units of $2e^2/h$), which serves as a strong experimental sign for the existence of TZMs. Our work not only clarifies the propagation nature of the THSs, but also realizes an array of 1D conducting channels on a 2D platform. This work provides an unprecedented way to engineering topological states intrinsic in TBG. 
\end{abstract}

\maketitle

\emph{Introduction.} 
Achieving one-dimensional (1D) conducting channels within a two-dimensional (2D) system presents a challenge. One approach involves utilizing edge states in topological systems, such as Chern insulators \cite{Haldane, SCZhang, Gordon, ZFang, CZChang, JCheckelsky} or $Z2$ topological insulators \cite{EJMele1, EJMele2, Bernevig, Molenkamp, Hasan, XLQi}. In these systems, the 1D channels are inherently localized at the boundaries, making them difficult to manipulate. Another approach involves introducing deliberate anisotropy \cite{Dirnberger2021Strain2D, Li2024ChernNumber, YFZhao2020QAHI}, such as  alternating stacks magnetic and undoped topological insulator layers \cite{YFZhao2020QAHI}. However, this method relies on artificial modifications rather than leveraging the material’s intrinsic properties. 

\begin{figure}
\includegraphics[width=8.4cm, clip=]{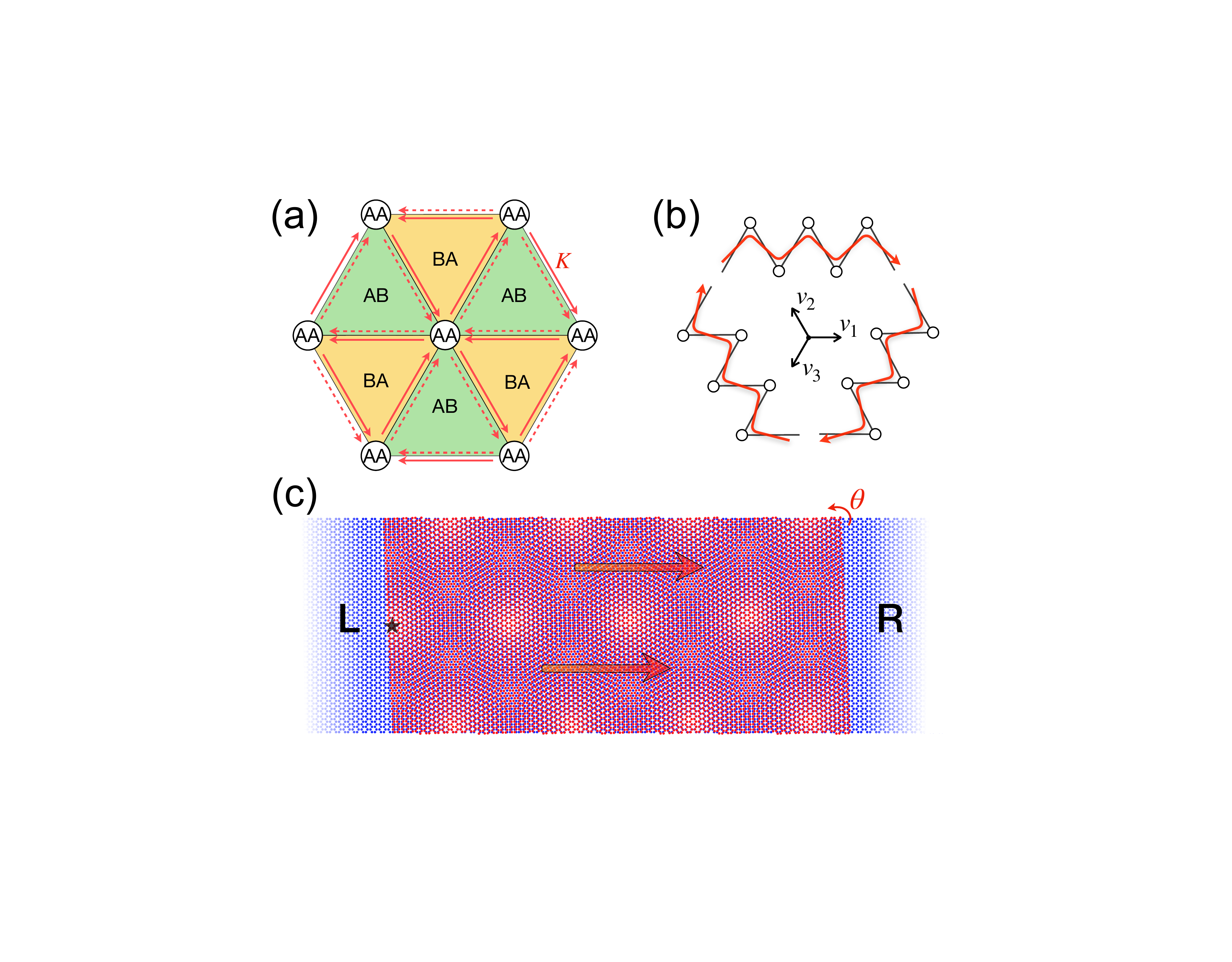}
\caption{(a) Stacking pattern and the 2D network model formed by the THSs in biased mTBG. For valley $K$, there are two chiral THSs depicted by the solid and dashes arrows between the AB/BA stacking regions. (b) Shematic diagrams for the black TZMs predicted by Ref. \cite{Tsim2020BiasTBG} and the red distorted TZMs. Circles denote the AA-stacking regions. The vectors ${\bm v}_{i}$ represent the velocities of the TZMs. (c) The two-terminal transport device adopted in our model, which can be decomposed into the left (right) lead made of the monolayer graphene with width $N$ (defined by the number of hexagons between two boundaries), and the central TBG flake with flat boundaries. The pentagram denotes the origin $O$ of the coordinate system which is also the twist center. The top-layer graphene is twisted anti-clockwise with angle $\theta$. The two giant arrows illustrate the longitudinal TZMs.  
}
\label{fig.Setup}
\end{figure}

Twisted bilayer graphene (TBG) has now emerged as a focal point in condensed matter physics, due to its unique flat band occuring at the magic angle ($\sim 1.05^{\circ}$) which provides an ideal platform for studying unconventional superconductivity and strong-correlation physics \cite{YCao2018SC, YCao2018CorrelatedInsulator, Stepanov2020SCMott, Wong2020ElectronicTransition, Andrei2020TBG, Sharpe2019Ferromagnetism, Lin2022FerromagnetismTBG, Zhang2020CorrelationTBG, Saito2021SubbandFerromagnetism, Serlin2020QAHEinTBG, Wu2021ChernInsulator}. Recently, minimally TBG with an applied interlayer bias has attracted considerable interest for the existence of the 1D topological helical states (THSs) at boundaries between the AB/BA stacking regions [Fig. \ref{fig.Setup} (a)] \cite{Walet2020, Prada2013Network, Efimkin2018Network, Beule2021mTBG, PWitting2023, Yoo2019TBG, Huang2018mTBG, Verbakel, Rickhaus2018Network, Xu2019Network, Tsim2020BiasTBG, Fleischmann2020mTBG, Beule2020ABTBG}. The origin of the THSs can be found from the difference of 2 of the valley Chern number per spin-valley at the domain walls \cite{IMartin2008, VaeziBG2013}. Till now, most works have believed that these THSs form the 2D triangular network \cite{Walet2020, Prada2013Network, Efimkin2018Network, Beule2021mTBG, PWitting2023, Yoo2019TBG, Huang2018mTBG, Verbakel, Rickhaus2018Network, Xu2019Network} [see Fig. \ref{fig.Setup} (a)] in which one THS is scattered into other three THSs at the AA stacking point, as can be inferred from the distribution of the local density of states (LDOSs) \cite{Walet2020, Huang2018mTBG, Verbakel}, and the experimental observations on the Aharonov-Bohm (AhB) oscillations in quantum transport \cite{Rickhaus2018Network, Xu2019Network}. Nevertheless, there exist opinions from the band-structure calculations \cite{Fleischmann2020mTBG, Tsim2020BiasTBG} and symmetry arguments from the one-parameter phenomenological scattering theory \cite{Beule2020ABTBG} that indicate that the THSs could self-construct the 1D topological zigzag modes (TZMs) [see Fig. \ref{fig.Setup}(b), the black curves], which propagate independently into three directions (${\bm v}_i, i =1, 2, 3$). The presence of TZMs indicates the potential for creating arrays of 1D conducting channels in minimally TBG. However, this potential has been hindered by quantum transport simulations in monolayer graphene systems with alternating staggered potentials \cite{THou2020, THou20201, You2022}, whose structure is topologically equivalent to biased TBG and supports the 2D network model.  

In this Letter, for the first time, we rigorously calculated the quantum transport through a minimally TBG system with interlayer bias, and support unambiguously the TZMs model. Our results are essentially distinct from Refs. \cite{THou2020, THou20201, You2022} because the realistic smooth transitions between the AA, AB, and BA stacking regions are captured. We show that, the THSs self-construct the 1D \emph{distorted} TZMs, bypassing any AA-stacking spot and propagating independently [see Fig. \ref{fig.Setup}(b), the red curves]. Importantly, these distorted TZMs induce a nearly quantized conductance plateau (NQCP) with a value close to 2 (in units of $2e^2/h$) when considering a long TBG nanoflake shown in Fig. \ref{fig.Setup}(c). The NQCP can be finely tuned close to 1 or 3 by a simple vertical shift of the twist center if one or three TZMs are embeded. A further calculation on the LDOSs of the nonequilibrium current validates the 1D and distorted nature of the TZMs.  Our work suggests the minimally TBG as a natural perfect platform for realizing arrays of 1D conducting channels, and provides an unprecedented way of engineering topological valley states in a biased TBG.

\emph{Model and methods.} 
We consider an untwisted graphene nanoribbon on the bottom with width $N$, and a twisted graphene flake on top. A flat boundary is adopted for the TBG region which can be achieved by placing a twisted rectangular graphene flake of length $L$ on top and then removing the un-overlapped regions through physical etching \cite{Xu2019Network, PalauTwistHetero2018, MahapatraQHTBG2022} or electrical gating \cite{DeschenesTBGEdgeStates2022, VriesTBGJosephson2021}. Its tight-binding Hamiltonian reads: \cite{Laissardiere2010Localization, Laissardiere2012NumericalStudies, Hou2024TBG}:
\begin{align}
\label{eq.Hamiltonian}
H=&   \sum_{ {\alpha}, i} (-1)^{\alpha}   \Delta | i, \alpha \rangle  \langle i, \alpha |  + \sum_{ \alpha, i, j } t^{\alpha \alpha}_{ij} |i, \alpha  \rangle  \langle j, \alpha |    \nonumber \\
& + \sum_{\alpha \neq \alpha'} \sum_{ i,j }   t^{\alpha \alpha'}_{ij} |i, \alpha \rangle  \langle j, \alpha' | ,
\end{align} 
where $\alpha = 1(2)$ denote the top(bottom) layer, and $| i, \alpha \rangle$ is the Wannier basis of the $p_z$ orbital at site $i$ with position ${\bm r}_{i}$. The first line of Eq. \ref{eq.Hamiltonian} denotes the intra-layer Hamiltonian of monolayer graphene with $\Delta$ the interlayer bias, and the second line represents the inter-layer hopping. The hopping integral $t^{\alpha \alpha'}_{ij}$ between site $i \in \alpha $ and $j \in \alpha' $ is determined by the Slater-Koster formula \cite{Slater1954HoppingFunction}:
\begin{align}
\label{eq.t_ij}
t^{\alpha \alpha'}_{ij} = & \left \{ V_{pp \pi}(r_{ij}) + 
(1 - \delta_{\alpha \alpha'} )  \chi^2  \left[ V_{pp \sigma} (r_{ij})  - V_{pp \pi}(r_{ij}) \right]  \right \}  \nonumber \\
&  \cdot \Theta \left(3 a - r^{\parallel}_{ij} \right),
\end{align}  
where $V_{pp \pi}$ and $V_{pp \sigma}$ are the intra(inter)-layer coupling strength which assume the position-dependent relations: $V_{pp \pi} (r_{ij}) = - \gamma_0 e^{  (a-r_{ij}) q_{\pi}/ {a}}$, and $V_{pp \sigma} (r_{ij}) =  \gamma_1 e^{ (a_I -r_{ij}) q_{\sigma}/ {a_I} }$. Here $a (a_I)$ is the intra(inter)-layer carbon-carbon atomic distance, and $\chi$ is defined as $\frac{ a_I }{r_{ij}}$ with $r_{ij}$ the atomic distance. The specific values of the parameters used here can be found in Ref. \cite{Hou2024TBG}. To reduce the computational cost, a hopping boundary of $3a$ has been set for the in-plane atomic distance $r^{\parallel}_{ij}$.

\begin{figure*}
\includegraphics[width=16cm, clip=]{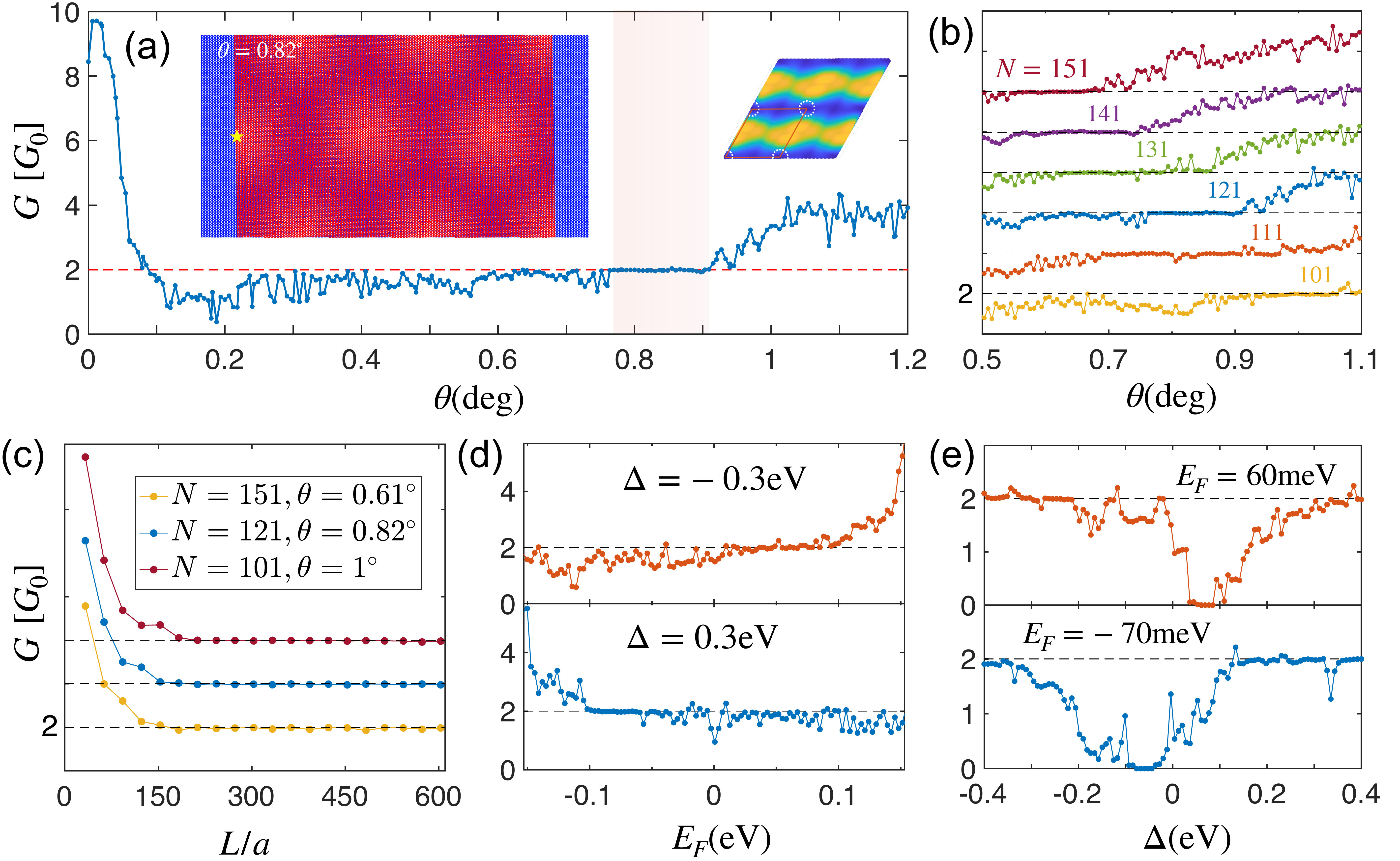}
\caption{Two-terminal conductance $G$ of the device with twist center fixed at the origin $O$. (a) $G$ versus the twist angle $\theta$. The red shaded region labels the NQCP of 2. Left inset: the atomic rigestory of the TBG. Right inset: distribution of the wavefunction for the longitudinal distorted TZM on the bottom layer, calculated at the commensurate angle $\theta_{com} \approx 0.8168^{\circ}$. The dashed circles denote the AA-stacking spots. (b) $G$ for different nanoribbon width $N$. (c) Length $L$ dependence of $G$. (d) $G$ versus the Fermi energy $E_F$ for $\Delta = \pm 0.3 {\rm eV}$ in the down (up) panel. (e) $G$ versus the interlayer bias $\Delta$ for $E_F = -70\ (60) {\rm meV}$ in the bottom (up) panel. For (a-c) $\Delta = 0.3 {\rm eV}$ and $E_F=-70 {\rm meV}$; for (a, d, e) $N=121$; for (a, b, d, e) $L = 303a$.   
}
\label{fig.Transport_CentralTwist}
\end{figure*}

\begin{figure*}
\includegraphics[width=11cm, clip=]{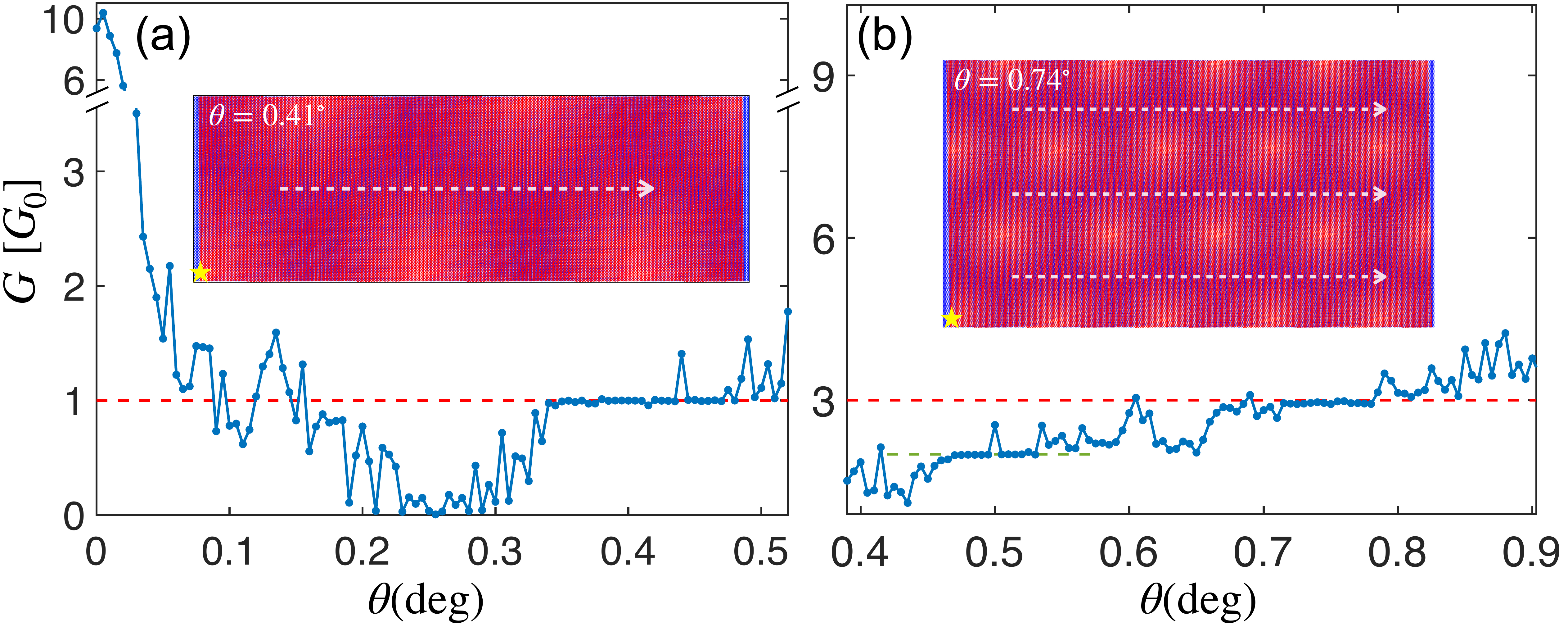}
\caption{$G$ versus $\theta$ with shifted twist center. In (a) $N=121$ and $E_F=-70$ meV. The twist center is at $(x_0, y_0) = (0, -56\sqrt{3}a)$. The inset shows the atomic registry at $\theta = 0.41^{\circ}$. In (b) $N=209$, and $E_F=-50$ meV. The twist center is at $(x_0, y_0) = (0, -100\sqrt{3}a)$, and the inset shows the atomic registry at $\theta = 0.74^{\circ}$. The green dashed line denote the value of 2. The white dashed arrows denote the propagation of the TZMs. In both (a) and (b) we set $L=603a$, and $\Delta = 0.3$ eV. }
\label{fig.ShiftedCenterTwist}
\end{figure*}

The two-terminal differential conductance $G$ through the central TBG flake at zero temperature is calculated using the non-equilibrium Green's function method \cite{Meir1992LandauerFormula, Jauho1994TransportResonant, Datta1995Mesoscopic}:
 \begin{align}
G(E_F) = \frac{2e^2}{h} {\rm Tr} [ {\bf \Gamma}_L {\bf G}^r_C {\bf \Gamma}_R {\bf G}^a_C],
\end{align}
where $E_F$ is the Fermi energy (defined relative to the charge neutral point of TBG with energy $E_{CNP} \approx 0.79$ eV), ${\bf \Gamma}_{L(R)} \equiv  i \left[ {\bf \Sigma}^r_{L(R)} - ( {\bf \Sigma}^r_{L(R)}) ^ {\dagger} \right]$ is the linewidth function for lead L(R) with the self-energy ${\bf \Sigma}^r_{L(R)}$ calculated iteratively from the monolayer graphene lead, and ${\bf G}^{r(a)}_C$ is the retarded(advanced) Green's function for the central region, satisfying: ${\bf G}^{r(a)}_C(E_F) = [(E_F \pm i 0^+){\bf I} - {\bf H}_C - {\bf \Sigma}^r_L - {\bf \Sigma}^r_R ] ^{-1}$, with ${\bf H}_C$ the Hamiltonian matrix of the central TBG region. In the following a conductance unit is defined: $G_0 \equiv 2e^2/h$. \\

\emph{Behavior of conductance.}
The number of longitudinal TZMs can be controlled by the twist center. For concreteness, we study the twist center at the origin $O$ [see the left inset of Fig. \ref{fig.Transport_CentralTwist}(a)] and show the conductance results in Fig. \ref{fig.Transport_CentralTwist}. Figure \ref{fig.Transport_CentralTwist}(a) plots $G$ versus the twist angle $\theta$ at the interlayer bias $\Delta = 300$ meV and the Fermi energy $E_F=-70$ meV. A sharp decrease of $G$ is observed for $\theta$ around zero, as the bilayer graphene is changed from AA-stacking to minimal twisting with Fermi velocity strongly surpressed \cite{Rakhmanov2012, SantosTBG2007}. For $0.76^{\circ} < \theta < 0.91^{\circ}$, a remarkable conductance pleateau is noticed (see the red shadow) with its value around 1.99, close to the quantized value of 2. Such emergence of the NQCP can be ascribed to the formation of the distorted TZMs from the band structure calculations, in which the distribution of the wavefunctions of the TZMs is shown in the right inset of Fig. \ref{fig.Transport_CentralTwist} (a)] [for more details, see Supplementary Information (S.I.)\cite{S.I.}]. Next we pick one twist angle from the plateau: $\theta = 0.82^{\circ}$, and plot its atomic registry in the left inset. As can be seen, there are three arrays of bright spots (AA-stacking spots) along the longitudinal transport direction, which flank two TZMs as shown by the giant arrows in Fig. \ref{fig.Setup}(c), explaining the emergence of NQCP. Further increasing the twist angle, the moir{\'e} pattern inside the TBG flake becomes narrower, allowing more TZMs encoded inside which always arise in pairs, and enhancing the conductance. However, no more conductance plateau is observed for larger twist angles as a consequence of the coupling between the longitudinal TZMs.

Figure \ref{fig.Transport_CentralTwist}(b) shows the transport behaviors under different width $N$ of the nanoribbon. Note that the quality of the NQCP becomes better with increasing $N$, indicating that the larger separation of the TZMs is, the better of the plateau. This also implies that higher NQCP with values 4, 6, 8, ... can be obtained if a much wider transport device is considered. Secondly, the NQCP moves left as $N$ increases, since to achieve two longitudinal TZMs embedded inside, a smaller twist angle $\theta$ is required at larger $N$. Figure \ref{fig.Transport_CentralTwist}(c) shows the length $L$ dependence of the conductance $G$. $G$ shows large values for very short $L$ due to quantum tunnelling. It then decreases quickly until saturating to the plateau by increasing $L$. At intermediate lengths (e.g., $L=123a$), except for the two longitudinally propagating TZMs [see ${\bm v}_1$ state in Fig. \ref{fig.Setup} (b)], there are also two more conducting modes with velocities $-{\bm v}_2$ and $-{\bm v}_3$ from $K'$ valley that mediate the transport. However, because of the vertical velocity component, these modes get filtered out as they reach the boundary when $L$ increases, preventing them from contributing to conductance \cite{Footnote1}.     

In Figs. \ref{fig.Transport_CentralTwist}(d) and \ref{fig.Transport_CentralTwist}(e), we investigate the Fermi energy $E_F$ and the interlayer bias $\Delta$ dependence of $G$, respectively, as a simulation for experimentally accessible scenarios. Here the size of the TBG flake and the twist angle are both fixed. The NQCP is observed in the hole (electron) excitation region for positive (negative) interlayer bias. This can be attributed to the layer polarization of the TZMs, whose wavefunctions concentrate on the bottom layer (see S.I. \cite{S.I.}), favoring a more transparent contact with the monolayer leads. Besides, in Fig. \ref{fig.Transport_CentralTwist}(d) the plateau in the hole side has a better quality than that in the electron side, as can be explained by the electron-hole asymmetry of the TZMs \cite{S.I.}. It is noted that for the positive interlayer bias, only the $K$-valley states get involved in transport from above analysis. For the reversed-bias, the valley index has been inverted to $K'$. As a result, a fully tunable and $100\%$ valley-polarized current is generated in our device.

To get other numbers of longitudinal TZMs, we shift the twist center to the lower boundary of the nano-flake and study how the plateau changes. A longer TBG flake  ($L = 603a$) has been chosen to fully filter out other non-longitudinal TZMs. In Fig. \ref{fig.ShiftedCenterTwist}(a), a prominent conductance plateau with its value around 0.999, close to the quantized value 1, is observed at $\theta > 0.34^{\circ}$. The atomic registry plotted in the inset is at $\theta = 0.41^{\circ}$, where the two AA-stacking arrays perfectly flank one TZM. In S. I. \cite{S.I.} we also examine the evolution of NQCP as the twist center is shifted. To achieve higher plateaus, we increase both the width of the nano-flake and the twist angle. As shown clearly in Fig. \ref{fig.ShiftedCenterTwist}(b), NQCP of 2 and 3 are observed. In the inset of Fig. \ref{fig.ShiftedCenterTwist}(b) the atomic registry at $\theta=0.74^{\circ}$ is plotted, which perfectly accounts for the plateau of 3. The coexistence of plateau of 1 and 3 indicate the 1D conducting channels must lie inside the bulk of the TBG, eliminating other explanations based on edge-states picture.

\begin{figure}
\includegraphics[width=8.4cm, clip=]{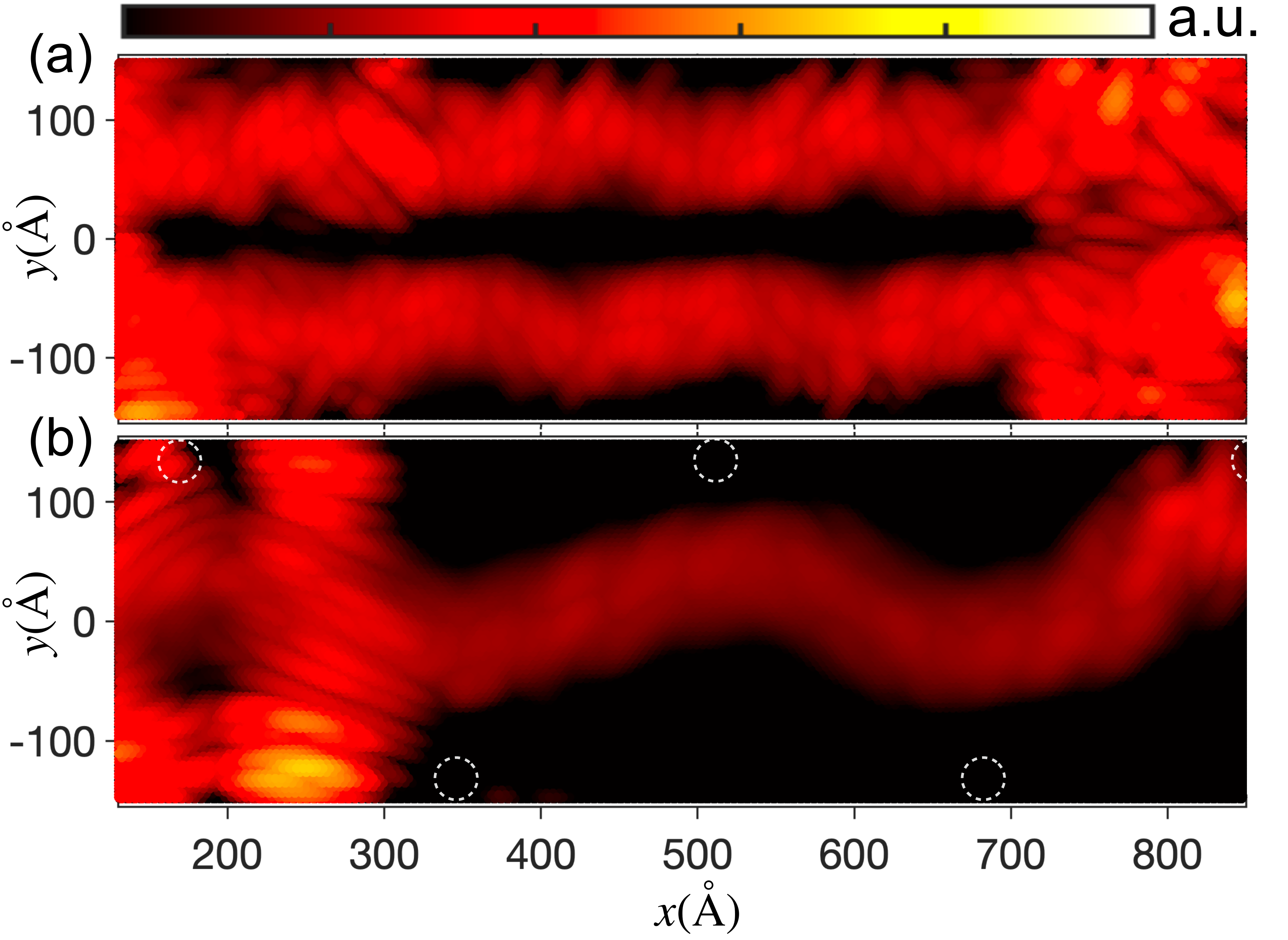}
\caption{Non-equilibrium LDOSs $\rho_{\rm NE}({\bm r}^b_i)$ on the bottom layer of the central TBG flake (here its logarithm is plotted). In (a) the twist center is in the origin $O$, and the twist angle is $\theta = 0.82^{\circ}$; In (b) the twist center is tuned close to the lower boundary of the flake with the atomic registry the same as the inset in Fig. 3(a). The dashes circles denote the AA stacking spots. Other parameters are the same as those of Fig. \ref{fig.ShiftedCenterTwist}(a). }
\label{fig.NEDOS}
\end{figure}

\emph{The Non-equilibrium LDOSs.}
To unambiguously demonstrate the role of TZMs in inducing the NQCP and their 1D nature, we present the LDOSs for the non-equilibrium current on the bottom layer: $\rho_{\rm NE} ({\bm r}^b_i)$ in Fig. \ref{fig.NEDOS}. This figure   details the propagation of local net current mediated by the TZMs (see S.I. \cite{S.I.} for the derivation on $\rho_{\rm NE}({\bm r}^b_i)$, and results for conductance plateau 3). The transport scenarios are chosen for the plateau of 2 with the twist center at origin $O$, and for the plateau of 1 with a shifted twist center, which has the same structure as the inset in Fig. \ref{fig.ShiftedCenterTwist}(a). Indeed, two distorted and well seperated TZMs stretching across the TBG flake are observed in Fig. \ref{fig.NEDOS}(a). The TZMs are repelled by the AA-stacking spots as can be found in Fig. \ref{fig.NEDOS}(b). These parallel TZMs never couple in real space. Here we want to comment that AhB oscillation should indeed happen in decive like Fig. \ref{fig.NEDOS}(a), but has a trivial origin from the enclosed trajectory formed by the two TZMs and the contacts. It also indicates that the experimentally observed oscillations may not account for the 2D network model \cite{Rickhaus2018Network, Xu2019Network}. The 1D channels demonstrated here are consistent with the Fabry-P{\'e}rot oscillation observed by Rickhaus {\it et al.} \cite{Rickhaus2018Network}.

\emph{Discussion and Conclusion.}
The NQCP demonstrates robustness against both the boundaries of the TBG flake and short-ranged point impurities, as shown in S.I. \cite{S.I.}. The robustness against the short-ranged impurities also proves the 1D nature of these TZMs, since electrons in spatially separated 1D channels experience a significantly lower scattering rate from such disorder. This also indicates that our protocol provides a practical scheme for the definitive experimental demonstration of the distorted TZMs. Furthermore, lattice reconstruction at minimal twist angles increases the AB/BA stacking regions and reduces domain boundaries \cite{Fleischmann2020mTBG}, thereby enhancing the quantized conductance plateau.  Additionally, all conducting channels are valley-polarized, highlighting their potential applications in valleytronics.

In conclusion, we studied quantum transport in a two-terminal mTBG device with interlayer bias and observed NQCP near 1, 2, and 3 for a wide range of twist angles, with spatially separated current distributions. These results confirm the 1D nature of the TZMs in the 2D twisted structure. Our findings resolve recent debates on the THSs in biased mTBG and demonstrate 1D conducting channels in a 2D system.

\emph{Acknowledgements.}
We thank Ya-Yun Hu, Jian-Peng Liu and Yu-Hang Li for fruitful discussions. This work is supported by the National Natural Science Foundation of China (Grants No. 12304070, No. 12350401), and the start-up grant of Nanjing Normal University (No. 184080H201B44).


\begin{thebibliography}{10}
\bibitem{Footnote1} Actually the plateau will finally collasplse for very long TBG since the zigzag mode is not strictly propagating along $x$ direction but has a small tilt angle ($\theta/2$) with respect to the $x$-axis.
\end{thebibliography}
\end{document}


\title{Supplementary Information for ``Arrays of one-dimensional conducting channels in minimally twisted bilayer graphene''}

\author{Zhe Hou}
\affiliation{School of Physics and Technology, Nanjing Normal University, Nanjing 210023, China.}
\author{Kai Yuan}
\affiliation{United Microelectronics Center Company Limited, Chongqing 400030, China.}
\author{Hua Jiang}
\affiliation{Interdisciplinary Center for Theoretical Physics and Information Sciences (ICTPIS), Fudan University, Shanghai 200433, China.}
\affiliation{Institute for Nanoelectronic Devices and Quantum Computing and State Key Laboratory of Surface Physics, Fudan University, Shanghai 200433, China.}

\maketitle
\tableofcontents

\newpage

\section{Band structure calculations on the TBG superlattice}

\begin{figure*}[h]
\includegraphics[width=16cm, clip=]{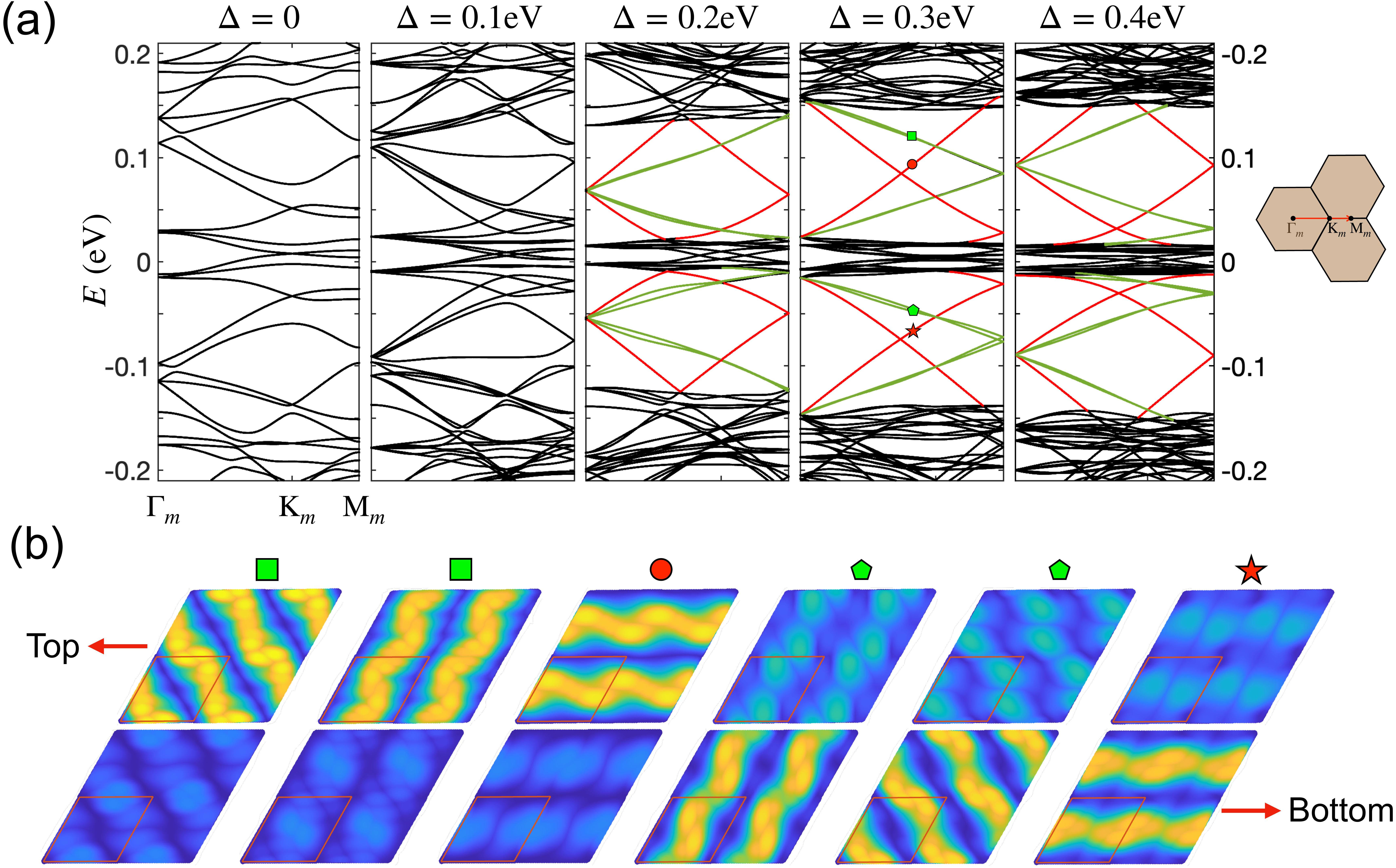}
\caption{ (a) Band structures of the TBG moir{\'e} superlattice along $\Gamma_m - {\rm K}_m - {\rm M}_m$ in the moir{\'e} Brillouin zone. Here the commensurate angle is about $\theta \approx 0.8168^{\circ}$, as can be characterized by two integer numbers $(40, 41)$. From the left to right the interlayer bias $\Delta$ is changed. For $\Delta \ge 0.2$ eV, two energy windows appear for the electron and hole excitation, which are spanned by the linearly dispersing modes (labelled by red and green colors). The rightmost panel shows the moir{\'e} Brillouin zone. (b) Distributions of the wavefunctions of the TZMs marked by colored spots in (a). Note that each green spot denote two states. The top(bottom) panels show the distribution on the top(bottom) layer. The red rhomboid denotes the unit moir{\'e} supercell. From the left to right the wavefunctions are shown in eigenenergy descending order.   }
\label{fig.BandStructure_Eigenstates}
\end{figure*}

\begin{figure*}[h]
\includegraphics[width=10.5cm, clip=]{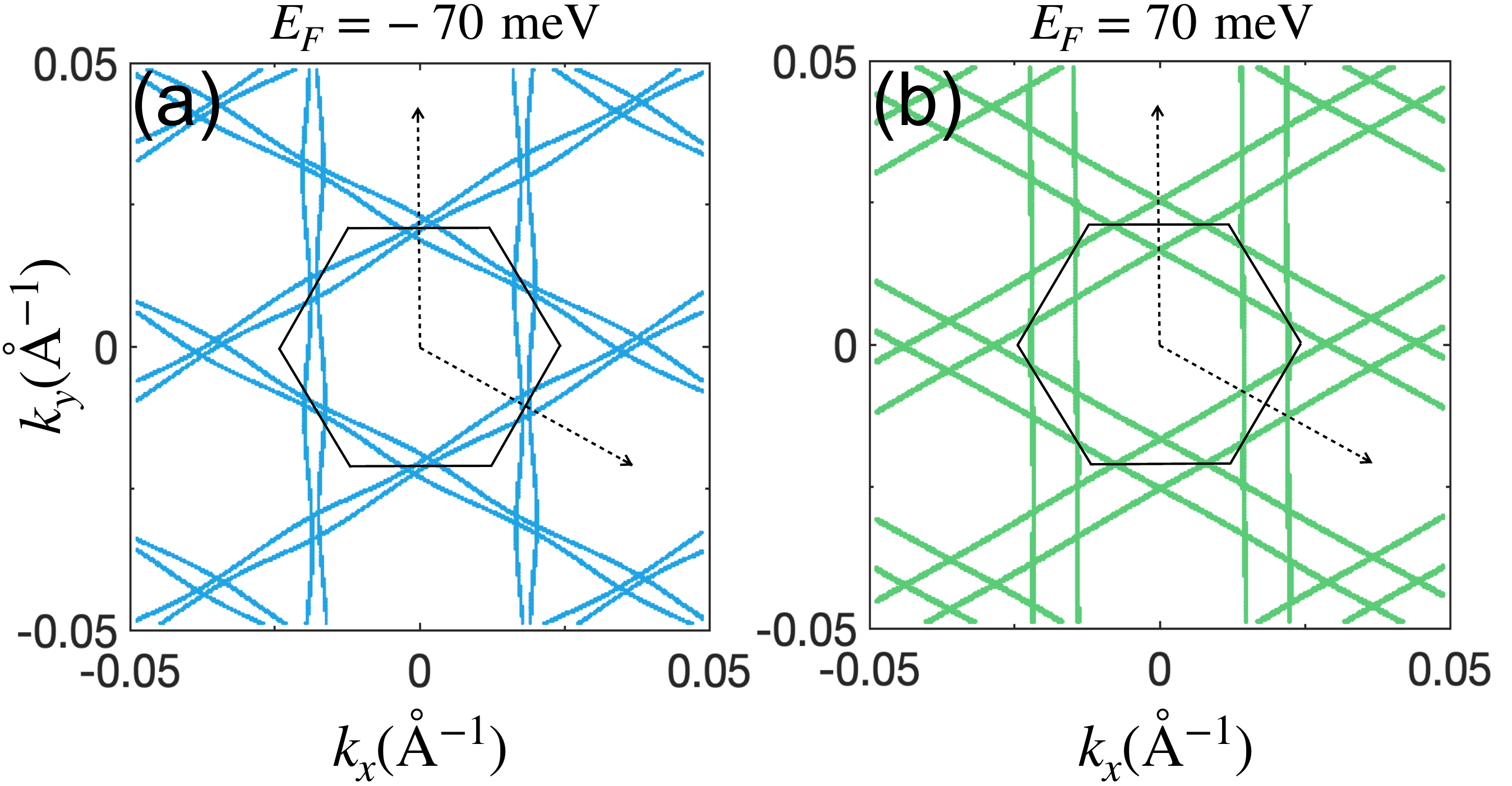}
\caption{Fermi surface of the TZMs in the moir{\'e} Brillouin zone for (a) $E_F=-70$ meV and (b) $E_F=70$ meV at $|\Delta| = 0.3$ eV. Here the black hexagon denotes the first moir{\'e} Brillouin zone, and the dashed arrows denote the two reciprocal lattice vectors of the superlattice. The commensurate twist angle is about $\theta \approx 0.8168^{\circ}$.
 }
\label{fig.BandStructure_FermiSurface}
\end{figure*}

To gain more insight into the distorted TZMs mediated conductance plateau, here we consider a periodic TBG moir{\'e} superlattice and plot the band structure as well as the eigenstates of the TZMs. The twist angle $\theta$ is chosen to be commensurate as characterized by two integer numbers $(m, n)$, following the relation: $\cos{\theta} = \frac{m^2 + 4 mn + n^2}{2(m^2 + mn + n^2)}$ \cite{Laissardiere2012NumericalStudies}. To investigate the TZMs at $ 0.82^{\circ}$ [see Fig. 2(a) in the main text], we choose $m=40, n=41$ whose commensurate angle is $\theta \approx 0.8168^{\circ}$. In Fig. \ref{fig.BandStructure_Eigenstates} (a) we plot the band structure along $\Gamma_m - {\rm K}_m - {\rm M}_m$ in the moir{\'e} Brillouin zone (see the rightmost panel). Note that the vector $\overrightarrow{\Gamma_m {\rm M}_m}$ has the same direction as the velocity ${\bm v}_1$ [see Fig. 1(b) in the main text], and $|\overrightarrow{\Gamma_m {\rm M}_m}|$ has the same length as the 1D Brillouin zone of a quasi-1D AA-stacking array (say, consider a quasi-1D TBG nanoribbon containing a single array of moir{\'e} pattern which has a strict periodicity along ${\bm v}_1$ direction). Here the interlayer bias is chosen from $\Delta = 0$ to $\Delta = 0.4$ eV. With the increase of $\Delta$, two energy windows on the electron and hole sides, spanned by the linearly dispersing states (labeled by red and green colors), can be observed. These linearly dispersing states are the TZMs, with propagating directions along $\pm {\bm v}_{i}$, $i=1, 2, 3$, and stretch across the whole moir{\'e} Brillouin zone. With increasing of  $\Delta$, the energy windows become wider, and more states are pushed to the charge-neutral point with their dispersion getting more flattened. These are the localized states on AA spots \cite{Tsim2020BiasTBG, Fleischmann2020mTBG}. The states propagating along $\pm {\bm v}_{2, 3}$ directions have velocity components projected along $\overrightarrow{\Gamma_m {\rm M}_m}$ which are half the value of the $\pm {\bm v}_1$ states. Using this property we can identify the $\pm {\bm v}_1$ states (shown in red) in Fig. \ref{fig.BandStructure_Eigenstates}(a) whose slopes are two times of the $\pm {\bm v}_{2, 3}$ states (shown in green). 

To check the eigenstates of the distorted TZMs, in Fig. \ref{fig.BandStructure_Eigenstates}(b) we plot the distribution of the wavefunctions on the moir{\'e} supercell by picking six different states in Fig. \ref{fig.BandStructure_Eigenstates}(a) with $\Delta = 0.3$ eV as marked by green and red spots (the two green states with close eigenenergies are marked by the same spot). Here the distribution of the wavefunctions on the top and bottom layers are shown separately in Fig. \ref{fig.BandStructure_Eigenstates}(b), with eigenenergies showing in descending order from left to right. Here the red rhomboid denotes the unit moir{\'e} supercell with its vertexes at the AA-stacking spots. As analysed, the four green states are indeed the TZMs with velocities $- {\bm v}_{2, 3}$, while the other two red states denote the longitudinally propagating TZMs with velocity ${\bm v}_1$. These states bypass all the AA-stacking spots, consistent with the non-equilibrium LDOSs calculated in Fig. 4 in the main text. More interestingly, the electron-like TZMs have wavefunctions dominant on the top layer, while the hole-like ones dominate on the bottom layer, indicating these TZMs are actually layer-polarized. This could explain why the conductance plateau for $\Delta >0$ can only be observed in the hole-like energy region [see   Figs. 1(d) and 1(e) of the main text]. 

To gain information of the wavefunctions about the negative bias case, we note that the positive and negative biased system can be related by a $C_2$ rotation along the $\bm{v}_1$ axis. This means that the band structures of the two reversed systems are identical, but all the TZMs have opposite layer-polarization. So for the negative bias case, the conductance plateau can only be observed in the electron-like energy region whose wavefunctions dominate on the bottom layer. One can also note that the wavefunction at the red pentagram on the bottom layer, is more separated than that of the red circle on the top layer. This shows again another sign of the particle-hole asymmetry of the TBG system besides the band structure, and explains why the conductance plateau observed for the positive bias (contributed by the red-pentagram state), is better than that observed for the negative bias (contributed by the reversed red-circle state) [see Fig. 2(d) in the main text].

Fig. \ref{fig.BandStructure_FermiSurface} shows the 2D Fermi surface of the TZMs at twist angle $\theta \approx 0.8168^{\circ}$. As can be seen, the whole Brillouin zone is filled with the parallel Fermi lines denoting the linearly dispersing TZMs.  \\

\section{Evolution of the NQCP from 2 to 1}
\begin{figure*}
\includegraphics[width=8cm, clip=]{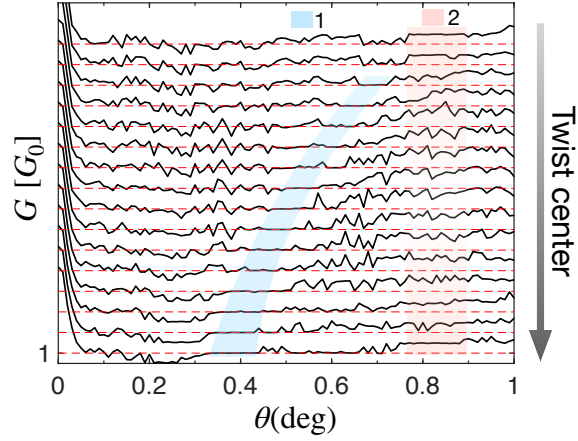}
\caption{Conductance curves as the twist center is moved downward with step $\delta y_0 = -4 \sqrt{3}a$. Here the twist centers for the topmost and downmost curves are set to origin $O$ and $(0, -60\sqrt{3}a)$, respectively. Each curve has been offset downward with a value of 2, and the dashed lines denote the position of value $1$. The parameters are: $N=121$, $L=603a$, $E_F=-70$ meV, and $\Delta = 0.3$ eV. The blue and red shaded regions label the NQCP of 1 and 2, respectively.} 
\label{fig.T_EvolutionFrom2To1}
\end{figure*}

\begin{figure*}
\includegraphics[width=16cm, clip=]{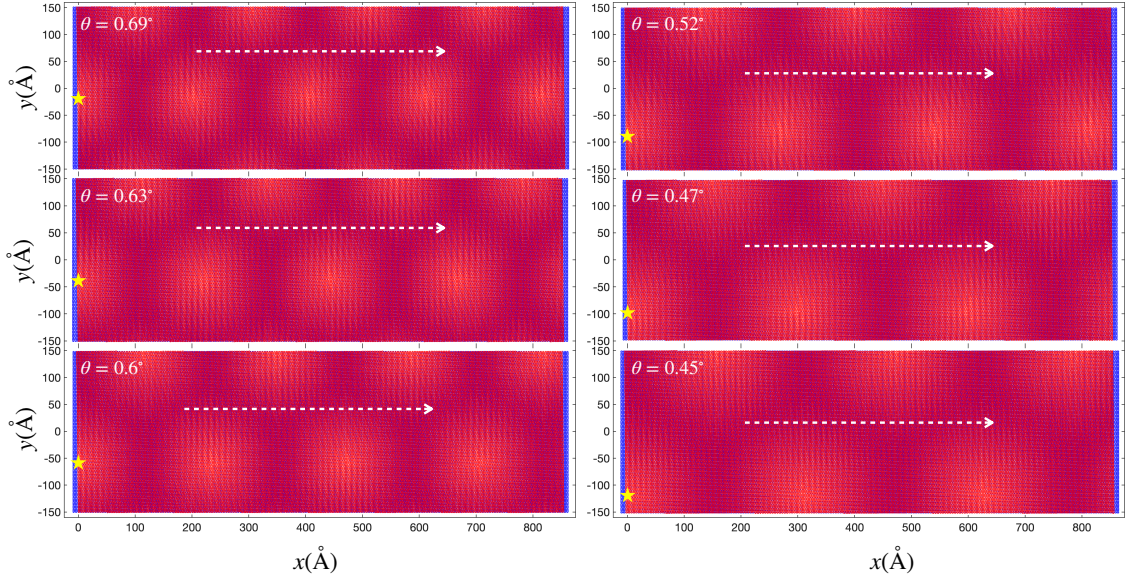}
\caption{Atomic registries of the central TBG nanoflake on the conductance plateau of 1 from Fig. \ref{fig.T_EvolutionFrom2To1}. Here the $y$-coordinates of the twist center (marked by the yellow pentagram) are chosen to be: $y_{0}=-8\sqrt{3}a, -16\sqrt{3}a, -24\sqrt{3}a, -32\sqrt{3}a, -40\sqrt{3}a$, and $-48\sqrt{3}a$, in $\theta$-descending order. The dashed arrows denote the propagation of the single TZM. }
\label{fig.AtomRegistry_EvolutionFrom2To1}
\end{figure*}

To see the evolution of the NQCP during the downward shift of the twist center, in Fig. \ref{fig.T_EvolutionFrom2To1} we plot $G$ versus $\theta$ by shifting the twist center from the origin ($y_{0} = 0$) to the bottom boundary of the nanoribbon ($y_{0} = -60\sqrt{3}a$), with step $\delta y_{0} = -4\sqrt{3}a$. Here each curve has been offset with a value 2 along with a red dashed eye-guide line denoting the conductance value of 1. The blue and red shaded regions denote the NQCP of 1 and 2, respectively. Note that the NQCP of 1 arises at $y_{0}=-8\sqrt{3}a$ for $0.66^{\circ} \leq \theta \leq 0.72^{\circ}$ as the formation of one TZM flanked by two arrays of AA-stacking spots (see the atomic registry in Fig. \ref{fig.AtomRegistry_EvolutionFrom2To1}). A smaller twist angle demands a larger vertical shift to achieve the perfect encoding of one TZM inside the flake, thus a leftward moving of the NQCP of 1 is observed. At the downmost boundary ($y_{0} = -60\sqrt{3}a$), the NQCP of 2 reappears as a result of the shift with a half of the period of the AA-stacking arrays. To show the correspondence between the moir{\'e} pattern and the plateau 1, in Fig. \ref{fig.AtomRegistry_EvolutionFrom2To1} we choose the $y$-coordinate of the twist center (marked by the yellow pentagram) from $y_{0}=-8\sqrt{3}a$ to $-48\sqrt{3}a$ on the conductance plateau of 1, and plot the atomic registries at fixed twist angles. As can be seen, for any stacking pattern, there is always one distorted TZM (depicted with dashes arrows) flanked by two parallel AA spots arrays, mediating the NQCP of 1. \\

\section{Derivation on the non-equilibrium LDOSs}
The non-equilibrium LDOSs is defined as the total LDOSs subtracting the LDOSs at equilibrium \cite{ZarboLocalCurrent2007}. To get its expression we start from the total LDOSs on site $p$ at time $t$ for steady current, which is defined as the expectation value of the particle number operator:
\begin{align}
\rho ({\bm r}_p) = & \langle  \hat{\psi}_{p }^\dagger (t) \hat{\psi}_{p} (t)  \rangle  
\equiv   - i G_{p, p}^< (t, t)    \\
=& \frac{-i}{2 \pi} \int_{-\infty}^{+\infty} d E G_{p,p}^<(E)   \nonumber \\
=& \frac{-i}{2 \pi} \int_{-\infty}^{+\infty} d E [G^r(E) \Sigma^<(E) G^a(E)]_{p,p}  \nonumber \\
=&  \frac{1}{2 \pi } \int_{-\infty}^{+\infty} d E \left[ G^r(E) [ f_L(E) \Gamma^L(E) +  f_R(E) \Gamma^R(E) ] G^a(E) \right]_{p,p} 
\end{align}
where $ \hat{\psi}_{p }^\dagger (t)$ [$\hat{\psi}_{p} (t)$] is the creation (annihilation) operator for site $p$ at time $t$, and $G_{p, p}^<$ is the correlation function (or lesser Green's function). On the second line a Fourier transformation into the energy representation has been performed, and on the third line we used the Keldysh equation: $G^<(E) = G^r(E) \Sigma^<(E) G^a(E)$ \cite{MeirLandauerFormula1992, JauhoTransport1994}, with $\Sigma^<(E) = i f_L(E) \Gamma^L(E) + i f_R(E) \Gamma^R(E)$. Here $f_{\alpha}(E)$ ($\alpha = L, R$) is the Fermi distribution function: $f_{\alpha}(E) = 1/ \{ \exp{[  (E- \mu_{\alpha} ) / k_B \mathcal{T} ]}  + 1 \} $ with $\mu_{\alpha}$ the chemical potential at lead $\alpha$, $k_B$ the Boltzman constant, and $\mathcal{T}$ the temperature. Throughout the paper we consider the zero temperature case: $\mathcal{T}=0$. So for $\mu_L > \mu_R$, the total LDOSs can be simplified into:
\begin{align}
\label{Eq.3}
\rho ({\bm r}_p) = &  \frac{1}{2\pi } \int_{-\infty}^{\mu_R} d E \left( G^r(E) [ \Gamma^L(E) +  \Gamma^R(E) ] G^a(E) \right)_{p,p}  +  \nonumber \\ 
&  \frac{1}{2\pi } \int_{\mu_R}^{\mu_L} d E \left[ G^r(E)  \Gamma^L(E)  G^a(E) \right]_{p,p} ,
\end{align}
where in Eq. (\ref{Eq.3}) the first line represents the LDOSs at equilibrium state: $\rho_{\rm E} ({\bm r}_p)$, which does not account for the net local current flow (in the presence of time-reversal symmetry) \cite{HJiang2009}; while the second line represents the non-equilibrium LDOSs $\rho_{\rm NE}({\bm r}_p)$ contributed from the net current flow from the left to the right terminal at non-equilibrium state (i.e., the steady state). 

For small chemical potential difference ($\mu_L \approx \mu_R \approx E_F$), we have for the second line of Eq. \ref{Eq.3}:
\begin{align}
G^r(E)  \Gamma^L(E)G^a(E) \approx  G^r(E_F)  \Gamma^L(E_F)G^a(E_F) ,
\end{align}
and the non-equilibrium LDOSs can be rewritten as:
\begin{align}
\rho_{\rm NE}({\bm r}_p) =  \frac{1}{2 \pi} (\mu_L - \mu_R) [G^r(E_F)  \Gamma^L(E_F)  G^a(E_F)]_{p,p} .
\end{align}\\

\section{Discussion on the conductance plateau of 3}

\begin{figure*}[h]
\includegraphics[width=16cm, clip=]{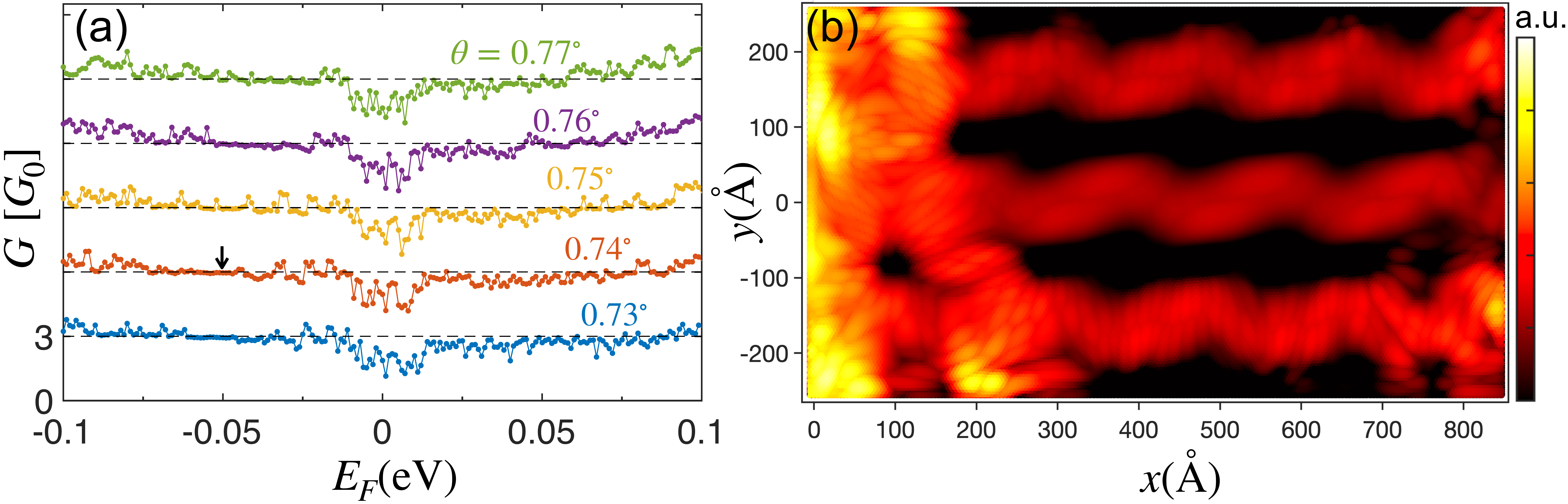}
\caption{(a): Fermi energy $E_F$ dependence of the conductance plateau at 3 under different twist angle $\theta$. Here the black dashed lines denote the conductance of 3, and each curve has been offset with a value of 3. (b): Logarithm of the non-equilibrium LDOSs $\rho_{NE}({\bf r}_i^b)$ plotted for $E_F=-50$ meV and $\theta = 0.74^{\circ}$ [marked by the short vertical arrow in (a)]. Here we set $N=209$, $L=603a$, the interlayer bias $\Delta$ = 0.3 eV, and the twist center $(x_0, y_0) = (0, -100\sqrt{3}a)$ $\approx (0, -246 \rm{\AA})$.   }
\label{fig.ConductancePlateauThree}
\end{figure*}

In Fig. \ref{fig.ConductancePlateauThree}(a) we show the Fermi energy dependence of the conductance $G$ around NQCP of 3. Here different twist angles $\theta$ supporting three TZMs are chosen. Since the interlayer bias is chosen to be $\Delta=$ 0.3 eV, NQCP of three can be seen in the hole-excitation region ($E_F < 0$). Next we pick one spot $E_F=-50$ meV from the conductance curve of $\theta=0.74^{\circ}$ (marked by the short vertical arrow) and plot its corresponding non-equilibrium LDOSs $\rho_{NE}({\bf r}_i^b)$ in Fig. \ref{fig.ConductancePlateauThree}(b). There three distorted TZMs can be seen clearly, validating the existence of the NQCP of 3.

\section{Boundary effect of the TBG nanoflake on conductance plateaus}
\begin{figure*}[h]
\includegraphics[width=13cm, clip=]{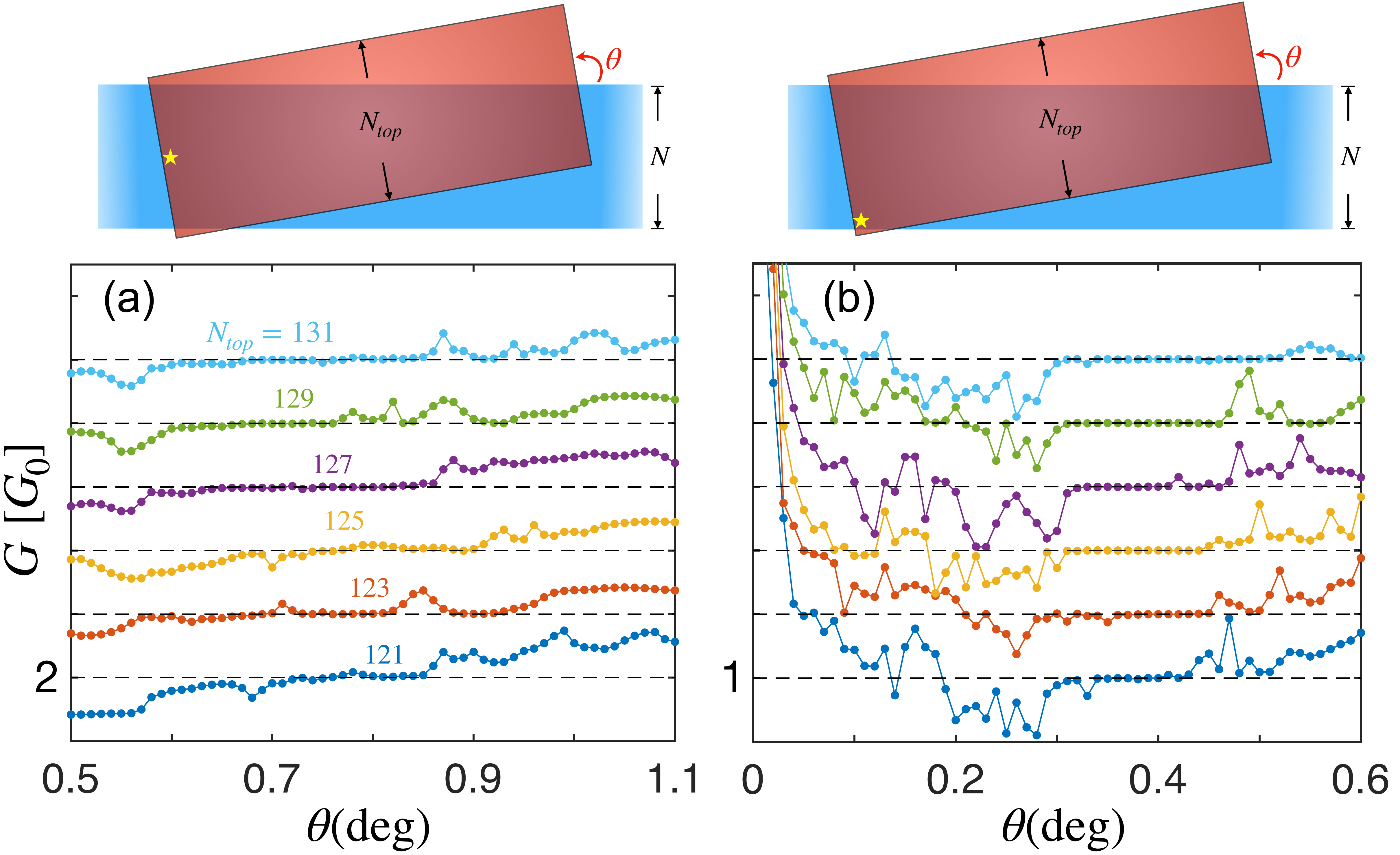}
\caption{Effect of nanoflake boundary on the NQCP. Here the different top layer width $N_{top}$ are chosen. In (a) the twist center is set at the origin and in (b) at the lower boundary with position $(x_0, y_0) = (0, -56\sqrt{3}a)$, as can be seen by the yellow pentagrams. Here (a) and (b) share the same legends. In (a) and (b) we set $N=121$, the interlayer bias $\Delta = 0.3$ eV, and the Fermi energy $E_F=-70 $ meV. For (a) the length of the central TBG region is $L=303a$ and for (b) $L=603a$.   }
\label{fig.BoundaryEffect}
\end{figure*}

In the main text, a flat boundary is used for the upper and lower boundaries of the central TBG nanoflake. This can be experimentally fabricated by twisting two rectangular graphene nanoflakes whose top layer has width $N_{top}$ much larger than the bottom one, and cutting-off the un-overlapped regions using physical etching \cite{PalauTwistHetero2018, MahapatraQHTBG2022} or electrical gating \cite{DeschenesTBGEdgeStates2022, VriesTBGJosephson2021}. To see whether the last step is necessary or not, we consider the twisted device whose top layer has almost the same width as the bottom one. The boundary of the central nanoflake is thus not flat anymore and is assumed to impact the conductance plateau. In the top panels of Fig. \ref{fig.BoundaryEffect}(a, b) we plot the schematic diagrams of the twisted nanoflakes, where the twist center (marked by the yellow pentagram) has been chosen at the origin and the lower boundary, respectively. We consider the parameters the same as those in Fig. 2(a) and Fig. 3(a) in the main text, and change the top layer width $N_{top}$ to see how the conductance plateau changes. For $N_{top}=N$, there are two un-overlaped monolayer regions parallel connecting with the TBG region. The plateau is affected and shows worse quality. Further increasing $N_{top}$, the upper un-overlapped monolayer region becomes larger while the lower un-overlapped region gets smaller. In this case, the NQCP of 2 whose twist center is at the origin, gets affected mostly. Instead, the NQCP of 1 whose twist center is at the lower boundary, is less affected. This can be attributed to the smaller twist angle at NQCP of 1, and the existence of single TZM, which does not couple with other TZMs. So, here we conclude that the NQCP of 2, mediated by two TZMs, are more sensitive to the nanoflake boundary than the NQCP of 1.
 \\

\section{Effect of a point impurity on the NQCP}
\begin{figure*}
\includegraphics[width=13cm, clip=]{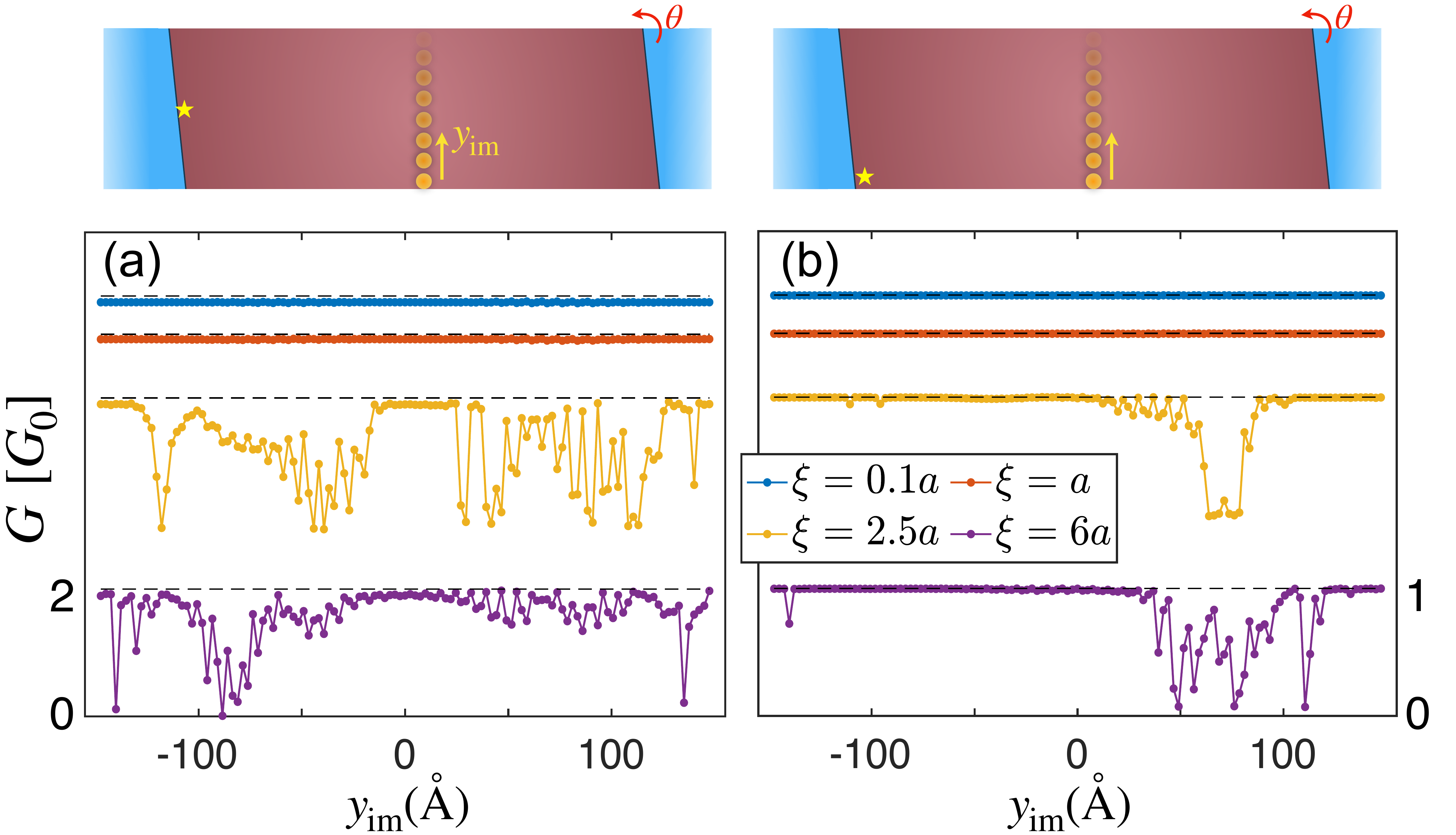}
\caption{Conductance under the influence of a point-impurity with correlation length $\xi$. (a) The twist angle is $\theta = 0.82^{\circ}$, and the twist center is set at the origin $O$. (b) The twist angle is $\theta=0.41^{\circ}$, and the twist center is set to $y_{0}=-56\sqrt{3}a$ at the lower boundary of the nanoribbon. Here The x-coordinate of the impurity is $X_{\rm im} \approx 500.05 {\AA}$, and the strength of the impurity is $V_{\rm im} = 2.5$ eV. The other parameters are the same as Fig. 4 of the main text.
}
\label{fig.ImpurityScattering}
\end{figure*}

To discuss a more experimentally relevant scenario where impurities should be accounted due to the effect of substrate, a point impurity is used here as a simulation. The point impurity has a Gaussian shape: $V_{im} ({\bm r}_i^b) = V_0 \exp{[ - | {\bm r}^b_i - {\bm r}^b_{\rm im}|^2 / (2 \xi^2) ] }$, where $V_0 = 2.5$ eV denotes the strength with the same amplitude as the nearest intra-layer hopping, ${\bm r}^b_i$ is the coordinate of the site $i$ at the bottom layer, and ${\bm r}^b_{\rm im} = (x_{\rm im}, y_{\rm im})$ is the position of the point impurity which is assumed to locate on the bottom layer. The correlation length of the impurity is denoted by $\chi$. We consider that the transport happens in the scenario of Fig. 4 of the main text and put the point impurity in the middle of the TBG flake by setting $x_{\rm im} \approx 500.05 {\AA}$. Fig. \ref{fig.ImpurityScattering} shows the conductance as a function of the impurity position $y_{im}$, where different correlation length $\xi$ has been adopted. We note that the conductance plateaus are immune to defect-like impurities ($\xi \leq a$), and is destroyed by impurities with larger $\xi$ only if the point impurity is placed on the way of the TZMs' propagation. The immunity to defect-like impurities can be understood by the diffraction effect of the TZMs, which have a broadening as large as $75 a$ as seen in Fig. 4 of the main text. This broadening allows the TZMs to remain unaffected by defect-like impurities. The results calculated here demonstrate the robustness of the TZMs against short-ranged impurities, as they can circumvent these impurities without scattering, providing further evidence of the 1D nature of TZMs in experimental measurements.  \\

